\documentclass{iopart}

\usepackage{setstack}
\usepackage{iopams}
\usepackage{amssymb}
\usepackage[T1]{fontenc}
\usepackage{graphics}

\newcommand{\vla}{\boldsymbol{\lambda}}
\newcommand{\kb}{k_{{\scriptscriptstyle B}}}
\newcommand{\kd}{{\kappa_{{\scriptscriptstyle D}}}}

\newcommand{\vk}{\mathbf{k}}
\newcommand{\vecr}{\mathbf{r}}
\newcommand{\vz}{\mathbf{0}}

\newcommand{\LN}{\Lambda_{\N}}
\newcommand{\LNk}{\LN (\vk; \vla)}
\newcommand{\LZ}{\Lambda_{\Z}}
\newcommand{\LZk}{\LZ (\vk; \vla)}
\newcommand{\ld}{\lambda^\dagger}

\newcommand{\jz}{j_{0}}

\newcommand{\Sd}{S_d}

\newcommand{\tu}{\tau  \upsilon}

\newcommand{\cO}{{\mathcal{O}}}

\newcommand{\Tc}{T_{c}}
\newcommand{\roc}{\rho_c}
\newcommand{\N}{{\scriptscriptstyle N}}
\newcommand{\Z}{{\scriptscriptstyle Z}}
\newcommand{\NN}{{\scriptscriptstyle N N}}
\newcommand{\ZZ}{{\scriptscriptstyle Z Z}}
\newcommand{\NZ}{{\scriptscriptstyle N Z}}

\newcommand{\SNN}{S_{\NN}}
\newcommand{\SZZ}{S_{\ZZ}}
\newcommand{\SNZ}{S_{\NZ}}

\newcommand{\DNs}{D_{{\N},\sigma}}
\newcommand{\DZ}{D_{\Z}}

\newcommand{\GNN}{G_{\NN}}
\newcommand{\GZZ}{G_{\ZZ}}
\newcommand{\GNZ}{G_{\NZ}}

\newcommand{\xiN}{\xi_{\N}}
\newcommand{\xiNi}{\xi_{\N, \infty}}
\newcommand{\xiZi}{\xi_{\Z, \infty}}
\newcommand{\xiZu}{\xi_{\Z, 1}}
\newcommand{\xiZd}{\xi_{\Z, 2}}
\newcommand{\xiZsp}{\xi_{\Z, \varsigma}^\psi}

\newcommand{\xiNu}{\xi_{{\N},1}}
\newcommand{\xiNs}{\xi_{{\N}\!,\sigma}}
\newcommand{\xiD}{\xi_{{\scriptscriptstyle D}}}
\newcommand{\xiDc}{\xi_{{\scriptscriptstyle D} , c}}
\newcommand{\fud}{\mbox{$\frac{1}{2}$}}

\newcommand{\RN}{R_{\N}}
\newcommand{\RNlr}{R_{\N}^{\LR}}
\newcommand{\Rflr}{R_{\varphi}^{\LR}}

\newcommand{\RZ}{R_{\Z}}
\newcommand{\RZlr}{R_{\Z}^{{\LR}}}

\newcommand{\vir}{\, ,}
\newcommand{\npt}{\, .}

\newcommand{\Io}{\mathcal{I}_{0}}

\newcommand{\nTr}{(T, \rho)}
\newcommand{\nTrc}{(T_c, \rho_c)}

\newcommand{\LR}{{\scriptscriptstyle{\mathcal{L}}}}
\newcommand{\sppd}{\mbox{when $\sigma <2$}}
\newcommand{\spgd}{\mbox{when $\sigma >2$}}
\newcommand{\ec}{\mathcal{E}_c}

\newcommand{\Phil}{\Phi^{{\LR}}}
\newcommand{\AF}{\mbox{{\textbf{AF}}}}
\newcommand{\df}{\delta_{\varphi}}
\newcommand{\dfBk}{\delta_{\varphi} B (\vk)}
\newcommand{\dl}{\dot{\lambda}}
\newcommand{\pTrp}{(T,\rho)}

\begin{document}

\letter{Charge and Density Fluctuations Lock Horns :\\
Ionic Criticality with Power-Law Forces}

\author{Jean-No\"el Aqua and Michael E. Fisher}

\address{Institute for Physical Science and Technology, University of 
Maryland, College Park, Maryland 20742, USA}

\eads{\mailto{jnaqua@glue.umd.edu}, \, \mailto{xpectnil@ipst.umd.edu}}

\begin{abstract}
How do charge and density fluctuations compete in ionic fluids near 
gas-liquid criticality when quantum mechanical effects play a role ? To gain 
some insight, long-range $\Phil_{\pm \, \pm} / r^{d+\sigma}$ interactions 
(with $\sigma>0$), that encompass van der Waals forces 
(when $\sigma \! = \! d \! = \! 3$), have been incorporated
in exactly soluble, $d$-dimensional 1:1 ionic spherical models with charges 
$\pm q_0$ and hard-core repulsions. In accord with previous work, when 
$d>\min \{\sigma, 2\}$ (and $q_0$ is not too large), the Coulomb interactions 
do not alter the ($q_0 \! = \! 0$) critical universality class that is 
characterized by density correlations at criticality decaying as 
$1/r^{d-2+\eta}$ with $\eta = \max \{ 0, 2\!-\!\sigma\}$. But screening is now 
algebraic, the charge-charge correlations decaying, in general, only as 
$1/r^{d+\sigma+4}$; thus $\sigma = 3$ faithfully mimics known 
\textit{non}critical 
$d=3$ quantal effects. But in the \textit{absence} of full ($+, -$) ion 
symmetry, density and charge fluctuations mix via a transparent mechanism:
then the screening \textit{at criticality} is \textit{weaker} by a 
factor $r^{4-2\eta}$. Furthermore, the otherwise valid Stillinger-Lovett
sum rule fails \textit{at} criticality whenever $\eta =0$ (as, e.g., 
when $\sigma>2$) although it remains valid if $\eta >0$ (as for $\sigma<2$ 
or in real $d \leq 3$ Ising-type systems). 
\end{abstract}



\nosections 

An ionic fluid, such as an electrolyte or a plasma, is characterized in 
thermal equilibrium by the screening of the long-range Coulomb interaction
potential, $z_\tau z_\upsilon q_0^2/ r^{d-2}$ (say, in $d>2$ dimensions) 
between ions of charges $z_\tau q_0$ and $z_\upsilon q_0$. 
Following Debye-H\"uckel theory 
\cite{HansenMacDo} one expects the charge-charge correlation 
function,\footnote{Our notations, which are fairly standard \cite{HansenMacDo}
are set
out explicitly in \cite{jn&mef04prl} which will be denoted {\AF}.}
$\GZZ (\vecr)$, to decay exponentially as $e^{-r/\xiZi}$ where,
as the overall ion density $\rho$ becomes small, the screening length 
$\xiZi \nTr$ should approach the Debye length 
$\xiD \equiv 1/\kd \varpropto \sqrt{T/q_0^2 \rho}$. 
More generally, the screening of an external charge in a conductor should be 
characterized by the Stillinger-Lovett sum rule \cite{HansenMacDo,stil&love68}
which does \textit{not} require \textit{exponential} screening \cite{mart88};  
this states that when 
$\vk \rightarrow \vz$, the charge structure factor, essentially the 
Fourier {transform\ddag} of $\GZZ (\vecr)$, behaves as 
\begin{equation}
  \label{sl}
  \SZZ (\vk) = 0 + \xiZu^2 k^2 - \xiZd^4 k^4 - \xiZsp k^\psi + \ldots 
  \quad \mbox{with} \quad \xiZu = \xiD \npt
\end{equation}
Note also that the vanishing leading term simply reflects the requirement 
of electroneutrality; if the screening is exponential only the further 
powers $k^{2 l}$ with $l = 1, 2, \ldots$ can arise; but, in general, which 
specific, nonanalytic powers appear and with what amplitudes is a matter of 
prime interest. 

The exponential screening of charge has been proven rigorously at low 
densities for classical systems \cite{Bryd&Fede80}; but sufficiently 
strong short-distance repulsions between oppositely charged ions are 
essential while any further non-Coulomb interaction potentials, say 
$\varphi_{\tau \upsilon} (\vecr)$ between ions of species $\tau$ and 
$\upsilon$, must be of short-range, decaying, e.g., exponentially fast when 
$r \rightarrow \infty$. But then, in allowing for quantum mechanics, one 
should first recognize that real polarizable ions also interact via 
fluctuating induced dipole-dipole or van der Waals forces that fall off 
only algebraically. Specifically, if for generality we consider integrable 
long-range potentials that decay as $\Phil_{\tau \upsilon}/ r^{d+\sigma}$ 
(so requiring $\sigma >0$) \cite{mef&ma72}, van der Waals forces may be 
described by $d\!=\!\sigma\!=\!3$. Then one may ask what will be 
the consequences 
of such long-range power-law forces for exponential screening and for 
the expansion (\ref{sl}). 

However, even in a point-ion model of a plasma, the quantum-mechanical 
fluctuations of position lead to effective particle-particle potentials, 
as manifest in the corresponding correlation functions, that decrease as 
$1/r^6$ (for $d=3$) \cite{Angel&Phil89,Cornu96II}, just as for polarizable 
ions (or neutral species). Furthermore, it is known that in quantum 
plasmas at low density, the screening of charge is no longer exponential:
indeed, $\GZZ (\vecr)$ has recently been shown to decay as 
$1/r^{10}$ \cite{Angel&Phil89,Cornu96II}. 

On approach to gas-liquid or liquid-liquid criticality, in an ionic system, 
the situation is further complicated because the density fluctuations 
become divergent and it is natural to ask if this should not seriously 
affect charge screening, the Stillinger-Lovett rule, etc. To be more 
explicit, in a short-range, \textit{nonionic} system, the density-density
correlation function $\GNN (\vecr)$ decays as $e^{-r/\xiNi}$ where, on the 
critical isochore $\rho = \roc$, the density correlation length $\xiNi \nTr$
diverges as $1/t^\nu$ with $t \equiv (T-\Tc)/\Tc$ and $\nu \geq \frac{1}{2}$; 
but \textit{at criticality},  one has
\begin{equation}
  \label{gnnc}
  \GNN^c (\vecr) \equiv D_{\NN}^c / r^{d-2+\eta} 
  \quad \mbox{when $r \rightarrow \infty$} \vir
\end{equation}
with $D_{\NN}^c$ finite and $\eta \geq 0$. Might not the divergence of 
$\xiNi$ couple in some way to the screening length $\xiZi$ in a 
corresponding ($\Phil_{\tau \upsilon} \equiv 0$) classical 
plasma and cause it to diverge at criticality ? Likewise, might not the 
slow decay of $\GNN (\vecr)$ at criticality change the $1/r^{10}$ quantal 
decay of $\GZZ (\vecr)$ that has been established at low densities ? 

Of course, the values of the critical exponents $\nu$ and $\eta$ depend on the 
critical universality class, and the influence of Coulomb couplings on 
criticality has been an important experimental and theoretical question ever 
since seemingly convincing observations on certain electrolytes suggested that
ionic criticality might realize a new or different type of critical behavior
[9-11].
While the experimental issues for 
electrolytes may now be regarded as largely settled --- in favor of no 
change in critical character --- the theoretical situation remains open. And 
neither experimentally nor theoretically has a clear picture of the charge 
correlations near criticality yet emerged. 

One question of particular relevance \cite{stel95} concerns the role of 
\textit{ion-symmetry}. Thus, the simplest theoretical description of a 
1:1 electrolyte is afforded by the so-called restricted primitive model or 
RPM, in which \textit{equisized} hard spheres carry charges $+q_0$ and 
$-q_0$. The RPM is precisely ion symmetric so that it is plausible that 
charge and density fluctuations will remain effectively independent even at 
criticality. Indeed, careful simulations \cite{luij&mef02,kim&mef04} have 
recently established that criticality in the $d=3$ RPM still exhibits 
Ising behavior as do simple, nonionic fluids. But in reality, $+$ and $-$ ions 
are never identical so that ion \textit{nonsymmetric} models are of 
especial interest. While simulations of 1:1, 2:1 and 3:1 hard sphere models 
with unequal diameters have been undertaken \cite{pana&mef02}, their 
critical behavior has not, as yet, been resolved. 

To address some of these problems, we have recently analyzed \cite{jn&mef04prl}
a class of exactly soluble \textit{d}-dimensional \textit{multicomponent}
spherical models in which particles of different species 
$\tau, \upsilon, \ldots$ reside on distinct but equivalent interlaced 
sublattices with nearest neighbor spacing $a$. 
Then, in a 1:1 `lattice electrolyte,' nearby ions of species $+$ and $-$
with charges $\pm q_0$ avoid collapse because opposite charges are never closer
than the minimal, intersublattice distance, say $a_0 \lesssim a$ 
\cite{jn&mef04prl}. 
In addition to the $\pm q_0^2 / r^{d-2}$ Coulomb potentials, 
\textit{short-range attractive potentials} $\varphi_{\pm \pm}^0 (\vecr)$ 
of magnitude, say, $\kb T_0$ are then introduced; on setting 
$q_0 \! = \! 0$, these suffice to yield standard spherical model critical 
behavior, with $\eta = 0$ etc., at $\Tc (q_0\!=\!0) \simeq T_0$ 
\cite{jn&mef04prl}.

Two main conclusions emerge from {\AF} \cite{jn&mef04prl}. First, the 
character of criticality remains unchanged when the charges are switched on
(provided $q_0$ is not too large). 
Second, ion symmetry plays a crucial role: thus in symmetric models, the 
charge screening length $\xiZi \nTr$ remains of order $\xiD$ even at $\nTrc$
although it gains singular corrections when $t \rightarrow 0$; 
but, on the contrary, for asymmetric fluids, $\xiZi$ \textit{diverges} on 
approach to criticality, precisely matching the density correlation length 
$\xiNi \sim 1/t^{\nu}$.  
Furthermore, the Stillinger-Lovett rule is then violated \textit{at}
criticality. Here we extend {\AF} by incorporating additional, 
long-range $\Phil_{\pm \pm}/r^{d+\sigma}$ ion-ion potentials in order to cast 
some light on the further role that quantum mechanical fluctuations might play.
(To avoid technical complications we suppose $\sigma \neq 2, 4, \cdots$ .)
As we report, the findings prove instructive. 

The necessary analysis follows closely the lines set out in {\AF}: accordingly
we focus on the principal results sketching only the basic technical points. 
(Further and fuller details will be presented elsewhere \cite{jn&mef04}.)
Thus Lagrange multipliers $(\lambda_+,\lambda_-) \equiv \vla \pTrp$ are 
introduced to satisfy the spherical model constraints 
$\langle s_\tau^2 \rangle \! = \! 1$ for $\tau \! = \! +,-$, where the 
$s_\tau ({\mathbf{R}})$ are the usual unbounded scalar spin variables 
[16-19].
The free energy then follows from the 
eigenvalues, 
$\LN (\vk; \vla)$ and $\LZ (\vk; \vla)$, of the interaction matrix 
${\mathbf{\Lambda}}$ with elements of the form 
$\Lambda_{\tau \upsilon} = \left[ \lambda_\tau \delta_{\tau \upsilon} - 
\tilde{\varphi}_{\tau \upsilon} (\vk)\right]$, where the 
$\tilde{\varphi}_{\tau \upsilon}$ are proportional to the Fourier transforms of
the total interaction potentials $\varphi_{\tau \upsilon} (\vecr)$. 

Now, as in {\AF}, the crucial result is that the $q_0^2/k^2$ Coulomb divergence
cancels out identically from the first eigenvalue which, at small wave 
numbers $k \equiv |\vk|$, then behaves as 
\begin{equation}
  \LNk  = \fud \kb T_0 \left[ \dl \pTrp + (\RN k)^2 + 
    (\RNlr k)^\sigma + \ldots \right] \vir  \label{dvlln}
\end{equation}
where $\dl$ ($\equiv \lambda/j_0$ in the notation of {\AF}) is found to vanish 
on the critical isochore near criticality like $t^\gamma$, while 
the finite, nonzero length 
$\RN (\vla; q_0)$ measures the \textit{range of the short range forces} 
[see {\AF} (25)]. The relative contribution of the 
\textit{long range interactions} to the density 
variation is embodied in the effective range $\RNlr$ \cite{jn&mef04}. 
From this form for $\LN$ one finds (for $q_0$ not too large) that the 
critical behavior is always of spherical model form with exponents 
$\beta = \fud$ and 
\begin{equation}
\label{expo}
  \eta (\sigma) = \max \{ 0 , \, 2-\sigma \} , \quad  \quad
  \gamma = 1 - \alpha = (2-\eta) \nu = 
    \frac{2- \eta (\sigma)}{d-2+\eta(\sigma)} \vir
\end{equation}
\cite{mef&ma72,joyc72,mef04} while $d>\min \{\sigma, 2\}$ 
is needed for $\Tc > 0$ 
[and we suppose $(\RNlr)^\sigma > 0$ when $\sigma\!<\!2 \,$ ].
For $\sigma\!>\!2$ these leading
exponents are, as well known, independent of $\sigma$ and the same as for 
short-range forces; but see also \cite{dant01}. However, new correction 
terms varying as 
$t^{\theta_\sigma}\!$ with $\theta_\sigma\!=\!|2\!-\!\sigma| \nu(\sigma)$ will 
dominate in all properties when $\sigma$ is close to $2$. 

The long range effects of the Coulomb forces appear only in the second 
eigenvalue which, for small $k$, varies as 
\begin{equation}
 \LZk  = \frac{\Sd}{4 a^d} \, \frac{q_0^2}{k^2} \,  
  \left[ 1 + (\RZ  k)^2 + (\RZlr k)^{2+\sigma} +  \ldots \, 
    \right]  \vir \label{dvllz}
\end{equation}
with $\Sd = 2 \pi^{d/2} / \Gamma(d/2)$. The net contribution of the long range 
forces is now represented by $\RZlr$ while $\RZ(\vla; \vk/k)$ is of order 
$a/\sqrt{\Io} \sim a / q_0$ where the \textit{ionicity} 
\begin{equation}
   \Io = q_0^2 / a^{d-2} \kb T_0 \vir 
\end{equation}
measures the overall strength of the Coulomb interactions near criticality. 
[It should be noted that the ellipses in (\ref{dvlln}) and (\ref{dvllz}) 
include
both further singular terms and analytic terms of order $k^4, k^6, \cdots$ .]

Finally, we may express the charge and density structure factors in the 
transparent form
\begin{equation}
  \label{decomp}
\eqalign{
  \frac{\SNN (\vk)}{\kb T / 4 \rho a^d} & = \frac{1-\dfBk}{\LN(\vk; \vla)} 
  + \frac{\dfBk}{\LZ (\vk; \vla)}, \\
&  \frac{\SZZ (\vk)}{\kb T / 4 \rho a^d} = \frac{\dfBk}{\LN (\vk; \vla)} 
  + \frac{1-\dfBk}{\LZ(\vk; \vla)} \vir }
\end{equation}
where the basic \textit{symmetry parameter} $\df$ ($=\ld/\jz$ in the notation 
of {\AF}) vanishes linearly with the deviation of the potentials 
$\varphi_{\tu}$ from precise (or effective) \textit{ion symmetry}. Clearly 
the density and charge fluctuations in ion symmetric models are 
completely uncoupled (at least in quadratic order): the critical density 
fluctuations are thus driven solely by $\LN (\vk)$ while charge 
screening is entirely controlled by $\LZ (\vk)$. But, charge and 
density fluctuations mix as soon as ion symmetry is lost: to what degree is 
determined by 
\begin{equation}
  \label{bk}
  B (\vk) = 4 k^4 a^4 \left[\df + (\Rflr k)^\sigma + \ldots \, 
  \right] \! / \Sd^2 \Io^2 \vir
\end{equation}
where $\Rflr$ measures the strength of the \textit{asymmetric}
parts of the long range forces (and hence vanishes with $\df$). 
A decomposition similar to (\ref{decomp}) holds for $\SNZ$: see {\AF}. 

The factor $k^4$ in (\ref{bk}) implies that the ``intrinsic'' charge 
fluctuations
contribute only weakly to $\SNN$. Consequently, except \textit{at} 
$(T_c,\roc)$, one has, neglecting analytic background terms of order $k^4$, 
\begin{equation}
  \label{snnk}
  \SNN (\vk) \varpropto 
  \frac{\chi_{{\scriptscriptstyle T}}}{1+ \xiNu^2 k^2
  +  \xiNs^{\sigma} k^{\sigma} + \ldots} 
  + \cO \! \left(\df k^{4+\sigma} \right) \vir
\end{equation}
where $\chi_{{\scriptscriptstyle T}} \sim 1/t^\gamma$ is the isothermal 
compressibility, while the new length scales are
\begin{equation}
  \label{xiNu}
  \xiNu \pTrp = \RN / \dl^{1/2} \quad \mbox{and} \quad 
  \xiNs \pTrp = \RNlr / \dl^{1/\sigma} \npt
\end{equation}
Since $\dl$ vanishes like $t^\gamma$ on approaching criticality, the 
\textit{density correlation length} 
may be identified as $\xiN \pTrp = \max \{ \xiNu, \xiNs \}$ in full 
accord with the exponent values (\ref{expo}). 

Likewise, the critical point decay (\ref{gnnc}) with $\eta = \max \{ 0 , \, 
2 - \sigma \}$ is readily verified. Away from criticality, however, matters 
are somewhat more subtle since, in general, no correlation function, 
$G_{\tu} (\vecr)$, can decay faster than the associated $1/r^{d+\sigma}$ 
power-law potentials \cite{benf&grub83}.  Indeed, the first nonanalyticity in 
$\SNN (\vk)$ yields the large-$r$ behavior \cite{ligh64} whence we find, 
for fixed $\pTrp \neq \nTrc$, 
\numparts
\begin{eqnarray}
  \label{gnnsppd}
  \GNN (\vecr; T, \rho)  \approx \frac{\DNs}{r^{d-2+\eta}}
  \left( \frac{\xiN}{r} \right)^{2\sigma} 
  \sim \frac{1}{r^{d+\sigma}} & \quad \sppd \vir \\
  \phantom{\GNN (\vecr; T, \rho) } \approx \frac{\DNs}{r^{d-2}} 
  \left( \frac{\xiN}{r} \right)^{2} \left( \frac{\xiNs}{r} \right)^{\sigma} 
  \sim \frac{1}{r^{d+\sigma}} & \quad \spgd \label{gnnspgd} \npt
\end{eqnarray}
\endnumparts
The appearance of the factor $( \xiNs / r)$ reflects the correction-to-scaling
exponent $\theta_\sigma = |2-\sigma| \nu (\sigma)$ identified above.

Now we may study the \textit{charge correlations} and ask, to start with, about
the \textit{noncritical behavior}. When ion symmetry pertains, i.e. $\df = 0$, 
the decomposition (\ref{decomp}) 
shows that $\SZZ$ depends only on $\LZ$; 
then (\ref{dvllz}) leads directly to the expansion (\ref{sl}) with, 
furthermore, full confirmation of the Stillinger-Lovett (SL) relation 
\cite{jn&mef04prl,smit88} (since 
$\kappa_{{\scriptscriptstyle D}}^2 = \Sd \rho q_0^2 / \kb T$). But, 
by virtue of the factor $k^4$ in (\ref{bk}), this remains true even when 
ion symmetry is absent. In addition, the coefficient of $k^4$ in (\ref{sl})
is given by 
\begin{equation}
  \label{xiZd}
  \xiZd^4 = \xiD^2 [ \RZ^2 - b_0 \df^2 \xiNs^{\, \sigma} \nTr ] \vir
\end{equation}
with $b_0 = \cO(a^{2-\sigma})$. Evidently, this moment of $\GZZ$ is 
unconstrained and if 
$\df \neq 0$ it will change sign when, \textit{driven} by the 
coupling to the \textit{density fluctuations} embodied in (\ref{decomp}), 
$\xiNs^{\, \sigma}$ diverges like $1/t^\gamma$ 
as $T \rightarrow \Tc$ on the critical isochore: see (\ref{xiNu}). 

Lastly, the leading nonanalytic term in $\SZZ (\vk)$ is given, in 
(\ref{sl}), by $\psi \! = \! 4+\sigma$ while its amplitude is 
\begin{equation}
  \label{xiZs}
  \xi_{\Z,\varsigma}^{4+\sigma} = \xiD^2 \left[ (R_{\Z}^{\LR})^{2+\sigma} 
    - b_1 \df (R_\varphi^{\LR} \xiNs)^\sigma + b_0 \df^2 \xiNs^{\, 2\sigma} 
    \right]
\end{equation}
where $b_1 = \cO (a^{2-\sigma})$. Clearly, this $k^{4+\sigma}$ term is 
present whether or not ion symmetry pertains; but if $\df \neq 0$ it may change
sign and its magnitude will diverge, like $1/t^{2\gamma}$, when 
$\nTr \rightarrow \nTrc$. More strikingly, however, the presence of this 
term ensures the \textit{destruction of exponential screening}; 
rather one finds \cite{ligh64}
\begin{equation}
  \label{gzzg}
  \GZZ (\vecr; T, \rho) \approx \frac{D_{\Z, \sigma}}{r^d}
  \left( \frac{\xi_{\Z, \varsigma}}{r} \right)^{4+\sigma} \sim 
  \frac{1}{r^{d+\sigma+4}} \npt 
\end{equation}
Thus near to \textit{or} 
far from criticality, long range forces always undermine the 
standard picture of Debye screening. Nevertheless, \textit{algebraic 
screening} remains in the charge correlations. Indeed, we may say that the 
$1/r^{d-2}$ Coulomb potential is screened by a factor $1/r^{6+\sigma}$ or, 
equivalently, that the long range $1/r^{d+\sigma}$ potential is 
screened --- owing to the requirements of ``local electroneutrality''---
by the factor $1/r^4$. This is a central result of our analysis and it is 
gratifying that on setting {$\sigma\!=\!d\!=\!3$} 
it reproduces the $1/r^{10}$ screening
previously found in a fully quantum mechanical analysis of point-charge plasmas
\cite{Angel&Phil89,Cornu96II}. 

In contrast to the loss of exponential screening, ion symmetry is paramount 
\textit{at criticality}. Thus we see from (\ref{dvllz}) and (\ref{decomp})
that all the results (\ref{xiZd})-(\ref{gzzg}) remain uniformly valid when 
$\nTr \rightarrow \nTrc$ provided $\df =0$, i.e., that \textit{ion symmetry
is valid}: one need only note that the amplitudes $\RZ^2$ and 
$D_{\Z, \sigma}(\xi_{\Z, \varsigma})^{4+\sigma}$ in (\ref{xiZd}) and 
(\ref{gzzg}) remain finite (and nonzero) at the critical point although 
the former will pick up a singular $t^{1-\alpha}$ correction (as in 
{\AF}: see \cite{jn&mef04}). 

On the other hand, for \textit{nonsymmetric fluids at criticality}, the
mixing of the charge and density fluctuations depends strongly on 
$\sigma$ or, more specifically, on $\eta(\sigma)$. Indeed, displaying only 
the leading singular terms, we find
\numparts
\label{szzk}
  \begin{eqnarray}
  \label{szzkc}
  \SZZ^c (\vk)  =  \xiDc^2 \, k^2  \left[ 1 + \df^2 (R_> k)^{2-\sigma}  
    +  \cO(k^2) 
  \right] & \mbox{\, for} \, \, \eta\!=\!2\!-\!\sigma >0 \vir \\
  \phantom{\SZZ^c (\vk)}  =  \xiDc^2 \, k^2 \left[ \ec - 
    \df^2 (R_< k)^{\sigma-2} 
  + \cO( k^2) 
  \right] & \mbox{\, for} \, \,\eta \!=\!0 < \sigma \!-\!2 \vir 
 \label{szzkcd}
\end{eqnarray}
\endnumparts
where $R_>^{2-\sigma}\!\!=\!\!\tilde{a}^2 (R_{\N}^{\LR})^{-\sigma}$, \, 
$R_<^{\sigma-2}\!\!=\!\!\tilde{a}^2 (R_{\N}^{\LR})^\sigma / R_{\N}^4$, \, 
$\ec\!=\!1 + \df^2 \tilde{a}^2 \!/\! R_{\N}^2 > 1$ with 
$\tilde{a}(q_0) = a \sqrt{2/\Sd \Io}$ : see {\AF} (29). 
One sees immediately that the Stillinger-Lovett sum rule \textit{remains 
valid} whenever $\sigma < 2$ or $\eta >0$. This contrasts with the results 
of {\AF} where, in the absence of long-range forces, one has $\eta = 0$
and the sum-rule is \textit{violated}, precisely the situation that prevails
here when $\sigma>2$; then the critical system may be regarded as an 
insulator or, at least, as an anomalous conductor!

From (15) 
we can now deduce the long-range behavior of the 
critical-point charge-charge correlation function, namely
 \numparts
  \begin{eqnarray}
  \label{GZZRc}
  \GZZ^c (\vecr)  \approx \frac{D^c_{\Z, \sigma} \, \df^2 \,  
    R_>^{2-\sigma}}{r^{d+4-\sigma}} \sim \frac{1}{r^{d-2+\eta}} \frac{1}{r^4}
    & \mbox{for} \quad \eta(\sigma) > 0 \vir \\
  \phantom{\GZZ^c (\vecr) }  \approx 
  \frac{D^c_{\Z, \sigma} \, \df^2 \,  
    R_<^{\sigma-2}}{r^{d+\sigma}} \sim \frac{1}{r^{d-2+\eta}} 
    \frac{1}{r^{2+\sigma}} \quad 
    & \mbox{for} \quad \eta = 0 < \sigma - 2 \npt 
  \end{eqnarray}
\endnumparts
As displayed, the results show that although $\GZZ^c(\vecr)$ is driven by the 
critical density fluctuations, with $\GNN^c \sim 1/r^{d-2+\eta}$, the charge 
correlations are screened relatively more strongly when $\sigma >2$ than 
for $\sigma <2$, when $\eta>0$. In light of the failure of the SL rule in the 
former case rather than the latter, this is, perhaps, paradoxical. On 
the other hand, one might equally conclude that the long-range $1/r^{d+\sigma}$
potentials are \textit{not} screened, even algebraically, when $\sigma >2$
(with $\eta\!=\!0$) whereas for $\sigma<2$ the long-range forces are actually 
screened by factors $1/r^{2\eta \, = \, 4-2\sigma}$. From that perspective the 
validity of the SL rule when $\eta >0$ seems more natural \cite{stel95}. 

\begin{table}
\caption{\label{resume} Long-distance behavior of charge correlations
$\GZZ(\vecr)$ at fixed $\nTr$. The density correlation length 
$\xiN$ diverges on the critical isochore $\rho = \roc$ as $1/t^\nu$ when 
$t = (T-\Tc)/\Tc \rightarrow 0$. The density correlations decay as 
$1/r^{d+\sigma}$ away from criticality, but \textit{at} criticality, since 
$\eta = \max \{ 0, \, 2\!-\!\sigma\}$, they decrease more slowly 
as $1/r^{d-\sigma}$ {\sppd}  and $1/r^{d-2}$ \spgd.}
\begin{tabular}{@{}cccc}
\br
  $\GZZ(\vecr)$ 
  & \textbf{ion symmetric} 
  & $ $ \hspace{45mm}
  & $ $ \hspace{-4.6cm} \textbf{nonsymmetric} \\  
  & $\df = 0$ 
  &
  & $ $ \hspace{-4.6cm}   $\df \neq 0 $ \\ 
  $ \nTr$    
  & 
  & $ $  \hspace{-7.5mm} $\sigma<2$ 
  & $ $  \hspace{-5mm} $\sigma>2$ \\
\mr
 = $\nTrc$, & $\sim 1/r^{d+\sigma+4}$ & 
 $\sim 1/r^{d+4-\sigma}$ & 
 $\sim 1/r^{d+\sigma}$ \rule{0mm}{4mm} \\
 $\neq \nTrc$, & $\sim 1/r^{d+\sigma+4}$ & 
 $\sim \left( \xiN/r \right)^{4-2\eta} / r^{d+4-\sigma}$ &  
 $\sim \left( \xiN/r \right)^{4} / r^{d+\sigma}$ \rule{0mm}{5mm}\\
\br
\end{tabular}
\end{table}

Finally, similar conclusions can be drawn about the charge-density structure 
factor: see {\AF}. \textit{Away from criticality} we find 
\begin{equation}
  \label{snzk}
  \SNZ (\vk) = \df \dl^{-1} \xiD^2 k^2 
  [ 1 - (\tilde{\xi}_{\N, \sigma} k)^\sigma +\ldots \, ] \vir
\end{equation}
where 
$\tilde{\xi}_{\N, \sigma}^\sigma = \xiNs^\sigma - (R_\varphi^{\LR})^\sigma/\df$
and one should recall from (\ref{xiNu}) that 
$|\xiNs^\sigma \nTr| \varpropto 1/\dl \nTr$ diverges like $1/t^\gamma$ while, 
by (\ref{bk}), $R_\varphi^{\LR}$ vanishes with $\df$. In real space the 
long-range decay can be written as 
  \begin{equation}
  \label{GNZ}
  \GNZ (\vecr; T, \rho)  \approx \frac{\df D_{\NZ}}{r^{d+|\sigma-2|}} 
  \left( \frac{\xiN}{r} \right)^{4\,-\,2\eta(\sigma)} \sim 
        \frac{1}{r^{d+\sigma+2}} \vir
  \end{equation}
so that the cross-correlation function is evidently screened by the factor 
$1/r^2$ relative to the long range potentials. Alternatively, one may say 
that the $1/r^{d-2}$ Coulomb interaction is screened by a factor 
$1/r^{4+\sigma}$. As regards the ``van der Waals case'' $d\!=\!\sigma\!=\!3$, 
we may 
mention that the charge density induced by an infinitesimal local external 
charge decays as $1/r^8$ in a fully quantal point-charge plasma
\cite{Angel&Phil89,Cornu96II}. 

\textit{At the critical point} itself, one finds that $\SNZ^c(\vz)$ vanishes 
identically if $\eta >0$ ($\sigma<2$) but that, as in the short-range case 
\cite{jn&mef04prl}, $\SNZ^c(\vz) = \df (\xiD/\RN)_c^2 \neq 0$, whenever  
$\eta = 0$ ($\sigma>2$). Moreover, when $r \rightarrow \infty$ one has 
$\GNZ^c(\vecr) \sim \df / r^{d+|\sigma-2|}$ so that the infection of the 
charge correlations by the density fluctuations again reduces the screening 
by a factor $r^{4-2\eta}$ that is greatest when $\eta = 0$: Compare with the 
results for $\GZZ (\vecr)$ as displayed in Table \ref{resume}.

In conclusion, we have analyzed the interplay between long-range density 
fluctuations, Coulomb interactions, and power-law $1/r^{d+\sigma}$ (e.g., 
van der Waals) interactions, away from, close to and \textit{at} 
criticality in ionic fluids, on the basis of two-component, $d$-dimensional  
spherical models with hard-core interspecies repulsions \cite{jn&mef04prl}. 
Throughout the fluid phase, including low-densities, 
the power-law forces destroy the usual Debye exponential screening. However, 
algebraic charge-charge and density-charge screening is still present: 
explicitly, although the 
density-density correlation function $\GNN (\vecr)$ decays no faster than the 
interaction potentials \cite{benf&grub83}, i.e., as $1/r^{d+\sigma}$, 
the charge-charge
correlation function, $\GZZ (\vecr)$, decays at large distances as 
$D_{\Z} \nTr / r^{d+\sigma+4}$. If one sets $\sigma\!=\!3$ to mimic the $1/r^6$
particle-particle interactions that arise in $d\!=\!3$ dimensions from 
quantal fluctuations \cite{Angel&Phil89,Cornu96II}, this result is, indeed, 
in accord with 
exact results for $G_{\ZZ} (\vecr)$ in a fully quantum-mechanical, 
point-charge plasma at low densities \cite{Angel&Phil89,Cornu96II}. 

In the critical region, the Coulomb interactions leave the universality class 
of the spherical models unchanged since, as in {\AF}, they still cancel 
out of the fluctuation factor that drives criticality. Nevertheless, radical 
changes arise in the charge-charge and charge-density fluctuations, 
whether the system is ion symmetric or, more realistically, nonsymmetric. The 
behavior is enforced by a general decomposition of the structure factors: see
(\ref{decomp}) and \cite{jn&mef04prl}. For ion symmetric fluids, the 
asymptotic amplitude $D_{\Z} \nTr$ is always finite 
and the Stillinger-Lovett (SL) sum rule is satisfied even at criticality. 
For asymmetric 
systems \textit{near} criticality, $\DZ$ is driven by the density 
fluctuations and hence diverges as $1/t^{2\gamma}$ when $\rho\!=\!\roc$ while 
the SL sum rule remains valid. \textit{At} criticality, however, the 
density correlations, decaying now as $1/r^{d-2+\eta}$, weaken the charge 
screening still further, more strongly when $\eta =0$ ($\sigma >2$) than if
$\eta >0$ ($\sigma<2$); see the summary in Table I. Finally, the SL sum rule, 
characteristic of 
conductors, is satisfied \textit{at} criticality when $\eta >0$, 
but is violated when $\eta =0$, a result that may well have validity beyond 
the ionic spherical models studied here \cite{stel95}. A further challenge is
to see how far similar results might be obtained for intrinsically quantal 
spherical models such as have been advanced in the past 
[18,23-25].

\ack
 Support from the 
 \textsc{NSF} (under grant CHE 03-01101) as well as assistance to 
 J.-N.A. from the French Ministry of Foreign Affairs under the 
 Lavoisier Fellowship program, is gratefully acknowledged. 


\end{document}